\documentclass[amssymb,amsmath,aps,showpacs,floatfix,nofootinbib,showpacs,12pt]{revtex4}
\usepackage{graphicx,epsfig}
\usepackage{color}
\def\beq{\begin{equation}}
\def\eeq{\end{equation}}
\def\be{\begin{equation}}
\def\ee{\end{equation}}
\def\bea{\begin{eqnarray}}
\def\eea{\end{eqnarray}}

\newcommand{\gsim}{\lower.7ex\hbox{$\;\stackrel{\textstyle>}{\sim}\;$}}
\newcommand{\lsim}{\lower.7ex\hbox{$\;\stackrel{\textstyle<}{\sim}\;$}}

\begin{document}


 \title{ Scalar resonance at 750 GeV as composite of heavy vector-like fermions}
 \author{ Wei Liao$^a$ and Han-qing Zheng$^{b,c}$}
 \affiliation{
  $^a$Institute of Modern Physics, School of Sciences \\
 East China University of Science and Technology,
 130 Meilong Road, Shanghai 200237, P. R. China %
 \vskip 0.2cm
\noindent{\small{$^b$ \it  Department of Physics and State Key
Laboratory of Nuclear Physics and Technology,
 Peking University, Beijing 100871, P. R. China}}\\
\noindent{\small{$^c$ \it   Collaborative Innovation Center of
Quantum Matter, Beijing 100871, P. R. China}}
}


\begin{abstract}
We study a model of scalars which includes both the SM Higgs and a scalar singlet as composites of 
heavy vector-like fermions. The vector-like fermions are bounded by the super-strong four-fermion 
interactions. The scalar singlet decays to SM vector bosons through loop of heavy vector-like fermions. 
We show that the surprisingly large production cross section of di-photon events at 750 GeV resonance 
and the odd decay properties can all be explained. This  model serves as a good model for both SM Higgs 
and a scalar resonance at 750 GeV.

\end{abstract}
 \maketitle

 {\bf Introduction}

CMS and ATLAS collaborations, with LHC  run-2 data of collected luminosity 2.6 fb$^{-1}$ and 3.2 fb$^{-1}$ respectively,
reported a possible resonance at around 750 GeV in di-photon channel~\cite{NewResonance}.
This resonance might correspond to the production of a spin 0 or  spin 2 particle with a mass around 750 GeV.
The production rate of these di-photon events is surprisingly large. If interpreted as  the production of
a particle of a mass of 750 GeV , the cross section of production at LHC with a center of mass energy of 13 TeV times the
branching ratio of di-photon decay is around a few fb.  Moreover, the resonance is interestingly only
reported in di-photon channel with no signal in WW, ZZ, Z$\gamma$ and $jj$ channels.
It's not surprising to see that this signal of the physics beyond the Standard Model(SM) gives rise to a challenge. 

While the reported resonance is still at a preliminary stage,
it's interesting to ask what does this resonance mean, if it's true, to the theory of (pseudo)scalars if it exists in nature.  This question is
interesting in particular by taking into account that a scalar, the Higgs particle of a mass 125 GeV, has
been discovered and it has properties very different from the reported 750 GeV resonance.
In this article, we will discuss the possibility of interpreting this 750 GeV resonance by a spin 0 particle and figure out the
implications to the theory of scalars.

In the next section, we will discuss several approaches interpreting this 750 GeV resonance by a spin 0 particle
and discuss why it is a great challenge to us.
Then we discuss a model which incorporates the SM Higgs
doublet and a singlet. In this model, the scalars are composites of heavy vector-like fermions(VLF)
which are bounded by strong four-fermions interactions.
It has properties similar to the BCS condensate, rather than the QCD condensate. In this sense, we call this
theory of scalars as  a BCS-like theory of scalars.  

 \vskip 0.5cm
 {\bf  Challenge to explain the 750 GeV resonance by a spin 0 particle}

The informations concerning this 750 GeV resonance 
are quite controversial.
If interpreted as the production of a spin 0 particle S,  it has peculiar features.
First, it has a very large production rate. If produced by gluon-gluon(gg) fusion, the cross section
to $\gamma \gamma$ channel can be estimated as
\bea
\sigma Br(\gamma\gamma)=\frac{1}{s \Gamma M_S} C_g \Gamma_{gg} \Gamma_{\gamma \gamma} ,
\label{Xsec1}
\eea
where $M_S$ and $\Gamma$ are the mass and the width of S, $C_g=2137$ for $\sqrt{s}=13$ TeV \cite{some1}
which is a number arising from the integral over parton distribution.  $\Gamma_{gg}$ and $\Gamma_{\gamma \gamma}$
are decay rates of S in $gg$ and $\gamma \gamma$ channels. In this article, we do not consider production
through other partons since they are harder to give a large cross section.

Second, not only there are no signals of this particle decaying to $WW$, $ZZ$,  Z$\gamma$, $Zh$, $hh$ or $gg$ pairs,
but also there are quite strong constraints on possible resonances through these channels.
The search of possible resonance through these channels, performed by CMS and ATLAS collaborations with LHC run-1 data, 
have found no signal~\cite{ATLAS1,CMS1,ATLAS2,ATLAS3,ATLAS4,ATLAS5,dijet}.
Using the reported upper bounds at 8 TeV and rescaling them to 13 TeV,
 one can get bounds on decay of S to these final states:
 \bea
& \frac{\Gamma_{Z\gamma}}{\Gamma_{\gamma \gamma}} \lsim 2, ~~
\frac{\Gamma_{ZZ}}{\Gamma_{\gamma \gamma}} \lsim 6,~~
\frac{\Gamma_{WW}}{\Gamma_{\gamma \gamma}} \lsim 20.\label{bound2} \\
&\frac{\Gamma_{Zh}}{\Gamma_{\gamma \gamma}} \lsim 10,~~
\frac{\Gamma_{hh}}{\Gamma_{\gamma \gamma}} \lsim 20.\label{bound3} \\
&\frac{\Gamma_{gg}}{\Gamma_{\gamma \gamma}} \lsim 1300.\label{bound4}
\eea
Apparently, tree level couplings of S to $VV$ are very problematic in view of these bounds.
A natural solution is that S couples to vector bosons all through loop, the same as for
couplings with $\gamma \gamma$ and gg. This means that S is more likely to be a SM singlet.

Third,  the width of the resonance seems to be quite large, i.e. $\sim 50$ GeV.  It's not easy to account for such
a large width with loop induced decay processes.
At the moment, the width analysis is very preliminary and analysis using narrow width approximation gives a
resonance with a confidence level as good as that using wide width approximation.  So there are no real measurement
of the width of this resonance and in this article we are not going to discuss too much this aspect of the reported resonance.

As a simple model of scalars, the two-Higgs doublet model(2HDM) has been considered a natural extension of the SM Higgs.
It can naturally accommodate, as the lightest scalar in the model, the 125 GeV Higgs observed in experiments.
Meanwhile, it has two heavy neutral Higgs in this model, one CP-even and one CP-odd, which potentially can account for the 750 GeV resonance.
But, there is a very serious problem concerning how to produce such a large production cross section in $\gamma \gamma$ channel
in this model of electroweak interaction~\cite{2HDM1}, not to mention how to fine-tune the model so that the heavy neutral Higgs can avoid 
tree-level decay into $WW$, $ZZ$ and $Z\gamma$ pairs.  

One way to enhance the production rate in 2HDM is to introduce extra particles, e.g. VLFs which is preferable to have a mass
half of the mass of S, i.e. $M_S/2$, so that the loop function reaches maximum.  It's apparent that these extra vector-like
fermions should be colored under the SM SU(3) group so that large enhancement of production through gluon fusion could be
generated.  On the other hand, vector-like quarks(VLQs) have been extensively searched by LHC experiments and
no signal of existence has been found up to a mass around 1 TeV~\cite{VLQ1,VLQ2}.
Since the Yukawa couplings of the neutral scalars in 2HDM with these VLQs are not related to their masses, 
unlike the case of top quark in SM,  the contributions to decays to  $\gamma \gamma$ and $gg$  by the VLQs are strongly suppressed
by the loop functions. A further assumption to save this scenario is to introduce many copies of VLQs. Needless to say, 
the price would be high and too many copies of VLFs would  have to be introduced. 
In short,  incorporating the 750 GeV scalar into a SM doublet makes the thing complicated and
does not give a satisfactory theory of scalars. 

Another approach to this 750 GeV resonance is to explain it as the remnant of a hidden QCD-like theory.
A virtue of a QCD-like model is that it can naturally give rise to a SM singlet.
One possible candidate of singlet is the glue-ball of the hidden sector 
when the VLFs in the hidden sector are heavy so that there are no pseudo-goldstone bosons in hidden sector. 
Another possible candidate is the hidden pion-like particle, if the masses of the hidden VLFs are light as in the case of QCD.

However,  these naive models have problems to understand the SM Higgs, a nice feature which makes simple models like 2HDM appealing.   
We can easily see that for the former case there is no hope to understand the SM Higgs as a glue-ball. 
In the latter case, three goldstone bosons generated from the color confining force can serve as the would-be goldstone bosons eaten 
by three weak gauge bosons, but $S$ cannot be produced by such type of theory since in QCD the $\sigma$ meson is very broad~\cite{zheng01}, 
not to mention producing the SM light Higgs.
In our point of view, there is no hope to reconcile 125 GeV resonance and 750 GeV resonance
into a common framework as originated from chiral symmetry breaking in the hidden sector.
In this sense, a naive QCD-like model of 750 GeV resonance, although might be interesting at first glance, is far from a theory of scalars.

There are also many other possible approaches and aspects to understand this 750 GeV resonance~\cite{some1,some1a, some2}. 
We are not going to discuss all of them in this article.  
In short, situations are similar in the sense that it's hard to understand this 750 GeV
resonance, in particular when considering it together with the SM Higgs in a universal theory.

\vskip 0.5cm
 {\bf  Higgs as heavy fermion condensate}

As discussed in the last section, a crucial hint from the experimental research is that VLQs, if exist,
should be quite heavy. If there is a theory of scalars in nature making use of heavy VLQs, this theory
has to give rise to low energy (pseudo)scalar degrees of freedom. Moreover, this mechanism, if it can explain the SM
Higgs and the 750 GeV resonance, should be different from a naive QCD-like theory described above. 
Indeed, there is a such mechanism in nature, i.e. the fermion condensate in BCS theory of superconductivity.
In BCS theory, the relevant fermion is electron which has a mass of MeV scale.
On the other hand,  the order parameter of the theory,  the fermion condensate which can be considered
arising from the dominating four-fermion interactions, give rise to physics of $10^{-3}$ eV scale, which is very far
away from the mass scale of electron.

There is no doubt that such a mechanism looks a bit peculiar for particle physics, since in vacuum there does not
seem to exist a Fermi surface  which is the crucial thing for this mechanism to work~\cite{Polchinski}.
Nevertheless,  this mechanism is still intriguing since it gives us an example of  non-decoupling heavy VLFs which
can inspire our thinking about theory of scalars,  and,  
after all, we don't really know what would happen at energy beyond the electroweak scale.
In this section, we present a model with fermion condensate  triggered by super-strong four-fermion interaction,
which also give low energy scalar degrees of freedom.
We will see that actually a model proposed 20 years ago~\cite{Zheng1} with a very simple upgrade by including a condensation in  the singlet channel via four-fermion interaction will overcome all the new challenges we are facing.

 We introduce two VLFs $\Psi^1$ and $\Psi^2$,  with  each $\Psi^i$ being a doublet $\Psi^i=(\psi^i_1,\psi^i_2)^T$.
 $\Psi^1$ and $\Psi^2$ form a doublet in the parity doublet space: $\Psi=(\Psi^1,\Psi^2)^T$.
 Then we introduce left and right fields $\Psi_l=P_l \Psi$ and $\Psi_r=P_r\Psi$ 
 in the parity doublet space where $P_{l,r}=(1\mp\sigma^2)/2$  and $\sigma^i$ is the Pauli matrix in the parity doublet space.
 $\Psi_l$ is charged under $SU(3)_C\times SU(2)_W\times U(1)_Y$.  The up and down components in $\Psi_l$ correspond to
 the isospin components of $SU(2)_W$ and they are all in fundamental representation of $SU(3)_C$.
 $\Psi_r$ is charged under $SU(3)_C \times U(1)_Y$. The up and down components in $\Psi_r$ correspond to
 isospin components of a global $SU(2)$ symmetry and they are both in fundamental representation of $SU(3)_C$. 
 We introduce the four-fermion interactions of $\Psi$
 \bea
 \Delta L= G_S  Y^2_S ({\bar \Psi} \Psi)^2
 +G_V Y^2_V\bigg[   ({\bar \Psi} \sigma^3 \Psi)^2+ ({\bar \Psi} \sigma^1 {\vec \tau} \Psi)^2  \bigg], \label{Lag1}
 \eea
 where  $\tau^i$  is the Pauli matrix in isospin space.  Color indices have been suppressed in (\ref{Lag1}).
 $Y_S$ and $Y_V$ are two dimensionless numbers.
  (\ref{Lag1}) is invariant under a $SU(2)\times SU(2)$ transformation
 in isospin space and the parity doublet space and consequently is invariant under a subgroup of it,  the $SU(2)_W$ group of the SM~\cite{Zheng1}.

 As long as $G_S$ or $G_V$ are super-large,  they can trigger condensate of fermion pairs.
 The condensate triggered by $G_S$ and $G_V$ can have some differences in energy scale.
 The condensate triggered by $G_S$ is a SM singlet  and we take this to be of higher energy scale, $\Lambda$.
 The condensate triggered by $G_V$ breaks the $SU(2)_W\times U(1)_Y$ symmetry of SM and we assume this second step condensation takes place a bit lower than $\Lambda$. Notice that $\Lambda$ can be substantially larger than the electroweak scale, $v\simeq 246$GeV.
 This is crucial in taking accurately the SM as the infrared limit of the dynamical model.

 Using techniques of bosonization, in a way how the Landau-Ginzburg theory is obtained in BCS theory of superconductivity, 
 we can eliminate,  at first step, the first four-fermion interaction term in (\ref{Lag1}) and
 get the effective interaction of a low energy bound state singlet ${ S}$ with fermion $\Psi$ at scale $\Lambda$:
 \bea
 \Delta L_1= -Y_S {S} {\bar \Psi} \Psi-\frac{1}{4G_S} S^2 \label{Lag2}.
 \eea
Higher order terms of $S$, such as $S^4$ and the kinematic term of $S$, can be obtained by running down to a lower scale~\cite{Zheng1,Zheng2}  .
The condensate of the fermion pair is understood as the development of  vacuum expectation value of scalar $S$
using this language.  
The bare fermion mass $M$ in Ref.~\cite{Zheng1,Zheng2} is generated in this way.
The second step condensate is described in details in Ref.~\cite{Zheng1} and a SM Higgs emerges. 
So this model is able to incorporate the SM Higgs and a singlet into a common framework.
The Yukawa coupling of SM Higgs with the SM fermions can be incorporated into this framework by including
in (\ref{Lag1})  some other four-fermion interactions.

As mentioned above, an immediate consequence of the development of a vacuum expectation value of ${ S}$, $\langle {S} \rangle =V_S$, is that
it gives a mass to VLQ $\Psi$
\bea
\delta  L_M=-M {\bar \Psi} \Psi -Y_S \tilde S{\bar \Psi} \Psi  ,~~M=Y_S V_S, \label{mass1}
\eea
where $\tilde S$ is the scalar  fluctuation which is a candidate of  heavy scalar particle and $Y_S$ is the
Yukawa coupling of $\tilde S$ to heavy fermion.
$M$ is the universal mass of heavy VLFs in this simple realization of the theory of scalars.
The condensate triggered by the $G_V$ coupling
has been discussed in the results in the emergence of the SM Higgs boson~\cite{Zheng1}.

Now we can see that scalar $\tilde S$  (herewith denoted as $S$), as a SM singlet, is a very good candidate for the 750 GeV resonance.
It does not have tree-level coupling to the SM particles and meanwhile can decay to gg,
$\gamma \gamma$ through the loop diagram of heavy fermions.  Note that there are no contributions of
vector boson loops in the decay of $S$ to $VV$. Very interestingly,  the contribution of decay of S to $VV$ by the
heavy VLFs does not decouple even if the fermion mass is much greater than $M_S$, similar
to the case of  top quark loop in SM to $h\to \gamma \gamma$ decay.  

To be specific, we take the $U(1)_Y$ charge of $\Psi_l$ as $Y$ and the $U(1)_Y$ charge of $\Psi_r$ as
$\frac{1}{2}\tau^3 +Y$, so that a Higgs doublet of right $U(1)_Y$ charge can be made from fermion pair~\cite{Zheng1}.
From these assignments we can get the coupling of $S$ to SM vector bosons and the decay rate to pairs of vector bosons:
\bea
&&\hspace{-0.7cm} \Gamma_{gg}  =\frac{\alpha_s^2 M_S^3 (2 N_\psi)^2}{144 \pi^3 V_S^2},  \label{Vgg}\\
&&\hspace{-0.7cm} \Gamma_{W W}= \frac{\alpha_2^2 M_S^3 N^2_c }{576 \pi^3 V_S^2},  \label{VWW}\\
&& \hspace{-0.7cm}\Gamma_{Z Z}= \frac{\alpha^2 M_S^3 N^2_c }{288 \pi^3 V_S^2}
\bigg(\frac{1}{2} \cot^2\theta_W+2 C_Y  \tan^2 \theta_W  \bigg)^2 \label{VZZ} \\
&&\hspace{-0.7cm} \Gamma_{Z \gamma}=\frac{\alpha^2 M_S^3 N^2_c  }{144 \pi^3 V_S^2}
\bigg( \frac{1}{2} \cot \theta_W-2 C_Y  \tan\theta_W \bigg)^2  \label{VZgamma} \\
&& \hspace{-0.7cm} \Gamma_{\gamma \gamma}= \frac{\alpha^2 M_S^3 N^2_c }{288 \pi^3 V_S^2}
\bigg(\frac{1}{2} + 2 C_Y  \bigg)^2  \label{Vgaga}
\eea
where we have taken the heavy fermion limit in loop function. 
$\alpha_2=\alpha/\sin^2\theta_W$, $\alpha$ the fine structure constant, $\theta_W$ the weak mixing angle, 
 $N_\psi=2$, $N_c=3$ and $C_Y=\frac{1}{4}+2 Y^2 $.

We can see that
\bea
\frac{\Gamma_{\gamma \gamma}}{\Gamma_{gg}}
=\frac{\alpha^2 N_c^2}{8\alpha_s^2 N_\Psi^2}(\frac{1}{2}+2 C_Y)^2 \label{ratio1}
\eea
It's about $1.7 \times 10^{-3}\times (\frac{1}{2}+2 C_Y)^2$  and it is quite safe with respect to the
constraint from the di-jet search. For example, for $Y=1/2$,  we have $C_Y=3/4$
and $\Gamma_{\gamma \gamma}/\Gamma_{gg}\approx 6.8\times 10^{-3}$ and
for $Y=0$,  we have $C_Y=1/4$ and $\Gamma_{\gamma \gamma}/\Gamma_{gg}\approx 1.7\times 10^{-3}$.
They all satisfy the bound in (\ref{bound4}). If without further assumption of the particle content,  
decay to gluon-gluon is the main decay channel and we can set $Br(S\to gg)\approx 1$.

The relative decay rate of other channels are
\bea
\frac{\Gamma_{Z \gamma}}{\Gamma_{\gamma \gamma}}
&&=\frac{2 (\cot \theta_W-4 C_Y  \tan\theta_W )^2}{(1 + 4 C_Y  )^2}, \\
\frac{\Gamma_{Z Z}}{\Gamma_{\gamma \gamma}}
&& = \frac{(\cot^2 \theta_W+4 C_Y  \tan^2\theta_W )^2}{(1 + 4 C_Y  )^2}, \\
\frac{\Gamma_{WW}}{\Gamma_{\gamma \gamma}}
&& =\frac{2}{\sin^4\theta_W (1 + 4 C_Y  )^2}
\eea
For $Y=1/2$,  we get
\bea
\frac{\Gamma_{Z \gamma}}{\Gamma_{\gamma \gamma}} \approx 0.004,~
\frac{\Gamma_{Z Z}}{\Gamma_{\gamma \gamma}} \approx 1.1,~
\frac{\Gamma_{WW}}{\Gamma_{\gamma \gamma}} \approx 2.3
\eea
For $Y=0$,  we get
\bea
\frac{\Gamma_{Z \gamma}}{\Gamma_{\gamma \gamma}} \approx 0.8,~
\frac{\Gamma_{Z Z}}{\Gamma_{\gamma \gamma}} \approx 3.2,~
\frac{\Gamma_{WW}}{\Gamma_{\gamma \gamma}} \approx 9.3
\eea
We can see that for both cases, (\ref{bound2}) and (\ref{bound4}) can all be safely satisfied. 
For larger $C_Y$,  decays to  $Z \gamma$, $Z Z$ and $WW$ channels 
are generally suppressed in comparison to $\gamma \gamma$ channel. This is due to the
effect of the coupling with the hypercharge field which can dominates the decay if $C_Y$ is increased.
If introducing $N_C$ copies $\Psi$, the decay rate can be increased by a factor of $N^2_C$.
We note that $C_Y$ and the di-photon decay rate can be increased by introducing vector-like leptons.
To just increase the total decay rate but not the di-photon decay rate, one can introduce
several copies of VLQs which are uncharged under $SU(2)_W\times U(1)_Y$. 
The decays of S to $Zh$ and $hh$ depend on the Yukawa couplings of SM Higgs to heavy VLFs and more details of the model.
It's not hard to see that the bound in (\ref{bound3}) can be satisfied by choosing the couplings of h with heavy fermions appropriately.
In the present article we do not elaborate on these decays and leave the detailed discussion on them to future works.

\begin{table}
\begin{tabular}{|c|c|c|c|c|c|c|c|c|c|}
 \hline
 ($N_C,Y$) & (2, 1/2)  & (3, 1/2) & (4,1/2) &(5,1/2)& (3, 0) & (4,0) & (5,0) & (6,0)\cr
 \hline
 $C_Y$ & 3/4 & 3/4 & 3/4 & 3/4 & 1/4 &1/4 & 1/4 & 1/4\cr
 \hline
 $V_S/M_S$ & 2 & 3 & 4 & 5 & 3/2 & 2 & 5/2 & 3\cr
 \hline
 \end{tabular}
 \caption{Parameters for the case with $N_C$ copies of parity doublet of VLQ. Approximate values of $V_S/M_S$, 
with which one can account for a fb scale di-photon production cross section, are given.  
$Y$ is assumed to be $1/2$ or $0$. As a reference, for $(N_C,Y)=(1,1/2)$ the value of $V_S/M_S$ is taken to be 1
in accord with Eq. (\ref{XSec}). } 
 \label{Table1}
 \end{table}

One can get for $Br(S\to gg)\approx 1$
\bea
\sigma Br(\gamma \gamma) \approx 3~\textrm{fb} ~\frac{ (1+4 C_Y)^2} {36} \frac{M_S^2}{V_S^2}.
\label{XSec}
\eea
We see that a cross section around fb can be produced by $M_S/V_S\sim 1 $ in this case. 
We note that for the case that there are a number of copies of parity doublet of VLQ or vector-like leptons, 
the di-photon production rate can be significantly enhanced. In this case, 
$V_S$ should be much larger than $M_S$ so that we can still get a fb scale di-photon production cross section.
For example, if $N_C$ copies of parity doublets of VLQ are introduced,  the production cross section in (\ref{XSec}) would be multiplied by $N^2_C$ and
$M_S^2/V_S^2$ should be suppressed by a factor around $1/N^2_C$ so that we can still achieve a fb scale
di-photon production cross section. Some examples for this case, in particular those with $Y=1/2$,  are shown in Table. \ref{Table1}.
We note again that in this case with many copies of parity doublets of VLQ,  
$V_S$ could be much larger than $M_S$ and the theory described here would be strongly coupled in a way very different from QCD in which $f_\pi$, 
the decay constant of $\pi$,  is close to the magnitude of the meson mass. 

We note that a RGE analysis on compositeness condition~\cite{BHL} for vector-like fermion condensation could 
be addressed in a way similar to that in~\cite{Zheng2}, by adding a new composite singlet scalar at 750 GeV. 
A new issue to be addressed is how to realize a two step breaking, i.e., firstly generate the mass of the vector-like fermions, 
and then the electroweak symmetry breaking.  A careful analysis on this aspect depends on detailed input of the model, 
e.g. the charge, number and the mass scale of the VLQs.  A detailed analysis of this aspect is out of the scope of
the present article and will be presented in the future. 

Nevertheless, some qualitative understandings can be obtained without detailed calculations. 
First, the effect of this singlet scalar on the running of  the SM couplings  is basically a two-loop effect and is small. 
Second,  the compositeness condition can say something about the fermionic content of the model and the width of the singlet scalar.
For example, if adding into the model some heavy VLQs which are uncharged under the electroweak group, 
the total width of S can be increased significantly while keeping the di-photon event rate the same.  However, too many
copies of these VLQs would destabilize the vacuum too fast at high energy and bring down the compositeness scale too close to
the electroweak scale so that causing problems in low energy phenomenology.  
Therefore,  for the simple model of scalars presented in the present article, 
the compositeness condition says that the fermionic content of the model can not be arbitrary. 
Third,  the RGE analysis of compositeness condition is although interesting and possible to say something,  e.g.,
about the fermionic content as discussed above,  but is hard to give a strong and quantitatively precise constraint on the low energy phenomenology. 
This is because, as actually pointed out long time ago~\cite{NonU},  
the low energy model of this type belongs to the same universality class of a renormalisable model. 
For the same reason, the four fermion interaction of the type presented in this article does not bring much new aspects on low energy phenomenology.

We further note that in principle, this hypothetical scalar can mix with the SM Higgs scalar and lead to interesting phenomenology. However, the
effect due to this mixing would be suppressed by $v^2/V_S^2$ where $v$ is the vacuum expectation value of the SM Higgs.
Moreover, the effect due to this mixing is not only model-dependent but also has a decoupling limit as $V_S$ tends to be large. 
We can see that one can tune the parameters to escape the possible experimental bound on this mixing.
Similarly, the signal of the VLQs in collider depends on the charges and the mass scale of VLQs, and whether they have mixings with
the SM quarks,  and is also very model-dependent. In the present article, we do
not study in detail these model-dependent aspects.  In general,  testing the model presented in the present
article in low energy phenomena is difficult and a direct test of the model
requires the discovery of parity doublet of VLQ and probably a collider in the future
with energy even higher than the LHC.

\vskip 0.5cm
{\bf Summary}

In summary, we have considered a model of scalars as composites of heavy vector-like fermions. 
The heavy vector-like fermions are bounded by the super-strong
four-fermion interactions and the condensate is triggered by the four-fermion interactions. This is similar
to the case of BCS theory. For this reason we call our theory of scalars as a BCS-like theory of scalars.
The scalar singlet obtained in the model can be naturally identified as the 750 GeV resonance. This singlet decays
to vector bosons all through loop of heavy vector-like fermions. We show that for reasonable parameters
we can reproduce the production cross section in di-photon channel. Interestingly, both the SM Higgs boson 
and 750 GeV resonance can be explained in this model as composites of heavy vector-like fermions and
a model originally proposed 20 years ago serves to be a good candidate to solve the new problem today.

\acknowledgments
This work of WL is supported by National Science Foundation of
 China (NSFC), grant No. 11135009, No. 11375065. The work of HQZ is supported by National Science Foundation of
 China (NSFC), grant No. 10925522.
 
 \vskip 0.5cm
{\it  Note added}: After the submission of the present article we got noticed articles \cite{Ellisetal,someothers}, that appear on arXiv
 a fews days before the present article, in which some scenarios with scalar singlet and heavy vector-like quarks 
 have also been discussed in explaining the 750 GeV resonance.
 The scenario of the present article, which is based on the model proposed in Refs. \cite{Zheng1,Zheng2},  
 has not been studied in those works.

\end{document}